\documentclass[conference]{IEEEtran}
\usepackage[utf8]{inputenc}
\usepackage[T1]{fontenc}

\usepackage{graphicx}
\usepackage{amsmath}
\usepackage{amsfonts}
\usepackage{cite}
\usepackage{subfigure}
\usepackage{url}
\usepackage{microtype}
\usepackage{float}
\usepackage{booktabs}
\usepackage{tikz}
\usepackage[capitalize,noabbrev]{cleveref}
\usepackage{units}

\begin{document}
%
\title{Evaluating the Electrification of Vehicle Fleets\\ Using the Veins Framework}

\author{
\IEEEauthorblockN{Sebastian Schellenberg, R\"udiger Berndt, Reinhard German, and David Eckhoff}
\IEEEauthorblockA{Computer Networks and Communication Systems, University of Erlangen-Nuremberg, Germany\\
Email: \{sebastian.schellenberg,ruediger.berndt,reinhard.german,david.eckhoff\}@fau.de}
}


%


\maketitle



%

\section{Introduction}

The simulative performance evaluation of future Intelligent
Transport Systems (ITSs) is a challenging task.
New mobility patterns, induced by car sharing or autonomous vehicles, the switch towards electric or hydrogen vehicles,
and the communication between vehicles (and infrastructure) are believed to change road traffic as we know it.
Investigating new applications and protocols, or finding answers to strategic questions---like the optimal placement of charging infrastructure---is often done by means of simulation.
The biggest challenge is the development of realistic models that are required for these simulations in order to produce meaningful results.

The case study discussed in this paper involves a company maintaining a vehicle fleet of one hundred vehicles.
In this article we will discuss how we extend and deploy the Veins framework~\cite{sommer2011bidirectionally}, which couples OMNeT++~\cite{varga2008overview} and SUMO~\cite{Krajzewicz2002}, to help in the process of electrifying this vehicle fleet, i.e., replacing combustion engine cars with electric vehicles to save money and lower CO$_2$ emissions.

%

\section{Modeling Electric Vehicles}

In order to study low level effects of vehicle fleet electrification an accurate model of electric vehicles is required.
The most important value during simulation is the current fill level of an electric vehicle's battery: the State Of Charge (SOC).

In \cite{batteryvtc2014fall}, we presented a lightweight battery model to be used in microscopic simulation.
We showed that the SOC can be calculated depending on the observed power flows.
The particular power components, i.e., the power to accelerate the car and to overcome street gradients, rolling and air resistance, are derived from the vehicle's kinematic.
By appointing fixed parameters like the vehicle's mass, all power flows can be determined with one input parameter only: the velocity.
This input is provided by the traffic simulator SUMO.
Furthermore, we include two additional components: the recuperation module (that calculates the regained power by transforming kinematic to electric energy during breaking and coasting) and 
the range extender module that represents an additional combustion engine that recharges the battery while driving.
The model's behavior has been validated using measurements from real test drives~\cite{batteryvtc2014fall}.


%
\section{Modeling the Infrastructure}

Vehicles powered by an electric engine need to be recharged when the battery is running out of capacity.
A range extender module generates electric energy, for example with a rotary engine, by using conventional fuel.
This energy is used to charge the battery and thus virtually extend its capacity, in particular while driving.

The recharging of an electric vehicle's battery takes considerably longer than refueling a conventional car.
The duration of the recharging process is determined by the transmission rate of electric power between charging station and vehicle, which,
in turn, is limited by the battery and the available charging infrastructure.
In order to comprehensively study the effects of electrification, our simulation model also incorporates the possibility to model charging infrastructure.

The infrastructure is composed of a number of charging stations configured in the following way.
Every station consists of a location, i.e., a point on the map where electric cars can park and charge their battery without blocking the road.
In our model, the location is reflected by an edge in the road network.
Secondly, each station has a number of charging slots including individual transmission rates per slot.
In our scenario, we assumed two different types of plugs: a common SchuKo plug, with a charging power of \unit[2300]{W}, and an IEC Typ 2 plug---standardized by the International Electrotechnical Commission (IEC)---with a charging power of \unit[3600]{W} (monophase and limited by max. charging power of reference car).
Finally, each station has a limit regarding the number of vehicles which charge at the same time.
In our scenario, it was possible to charge up to two vehicles per charging station simultaneously.

A charging manager has been implemented to control the different charging stations and the recharging processes.
If more cars want to charge than possible, they are queued using a first-in first-out strategy.
Our vehicles can independently decide to wait in line or drive to another station depending on the current SOC and the distance to the next charging station.

%
\section{Controlling the Scenario}

The simulation framework consists of two bidirectionally coupled simulators:
SUMO (Simulation of Urban Traffic), reflecting the movement of vehicles on the city map and OMNeT++, scheduling the usage of electric vehicles and controlling the charging infrastructure.

In order to increase the level of realism in the simulation, other cars can be incorporated within the SUMO network to reflect realistic traffic densities with respect to a specific time of day.
To keep simulation run times at a reasonable level, these vehicles will only exist on the SUMO side of the simulation and will not be represented by nodes in OMNeT++.

\begin{figure}[!h]
\centering
\includegraphics[width=0.4\textwidth]{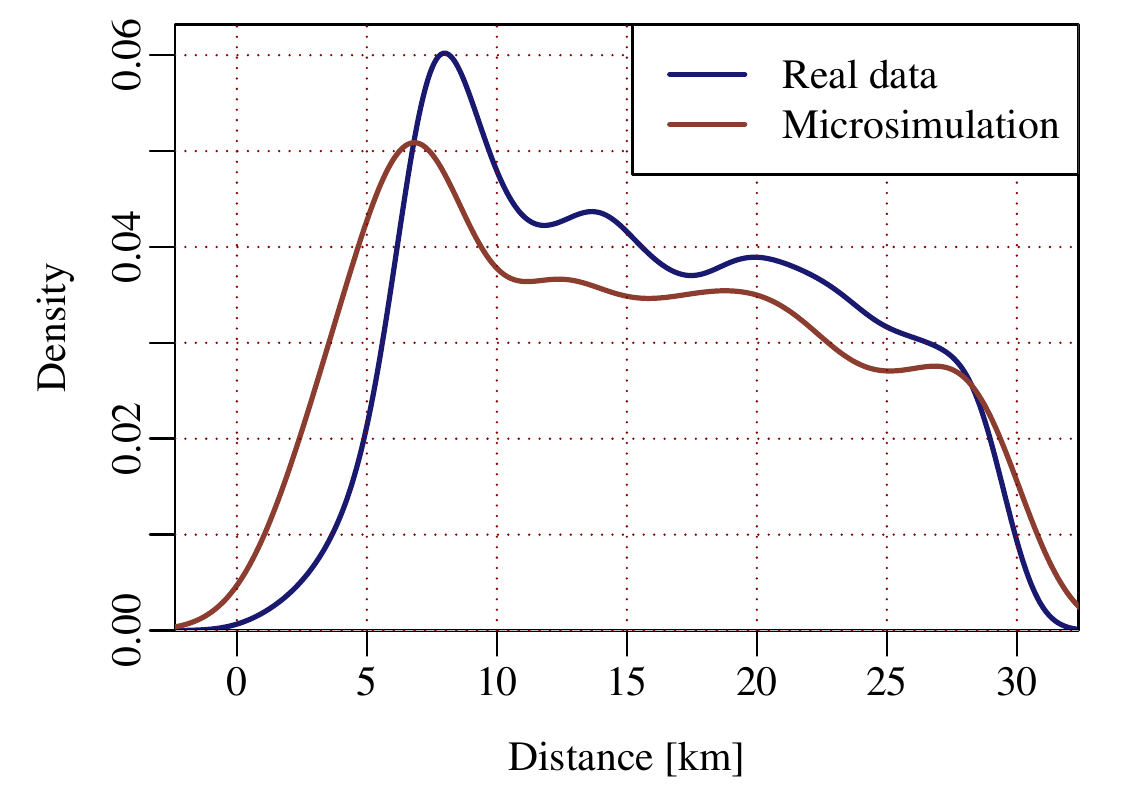}
\caption{Comparison between real-world and simulated driving distances}
\label{fig:hist}
\end{figure}

The schedule of the electric vehicles' usages is derived from statistical data of a real-world vehicle fleet (N-ERGIE Aktiengesellschaft) and used as input for the simulation.
An extension to the Traffic Control Interface (TraCI), already implemented in Veins and SUMO, allows OMNeT++ to generate vehicles according to the collected empiric data.
First, we choose a point on the map with an air-line distance according to a distribution obtained from the empiric data.
Then the nearest road to this point is used as the destination for the vehicle's route.
Naturally, the real driving distance differs the from air-line distance, but still satisfies the requirements for our simulation, as illustrated in \cref{fig:hist}.

\begin{figure}[h]
\centering
\includegraphics[width=0.4\textwidth]{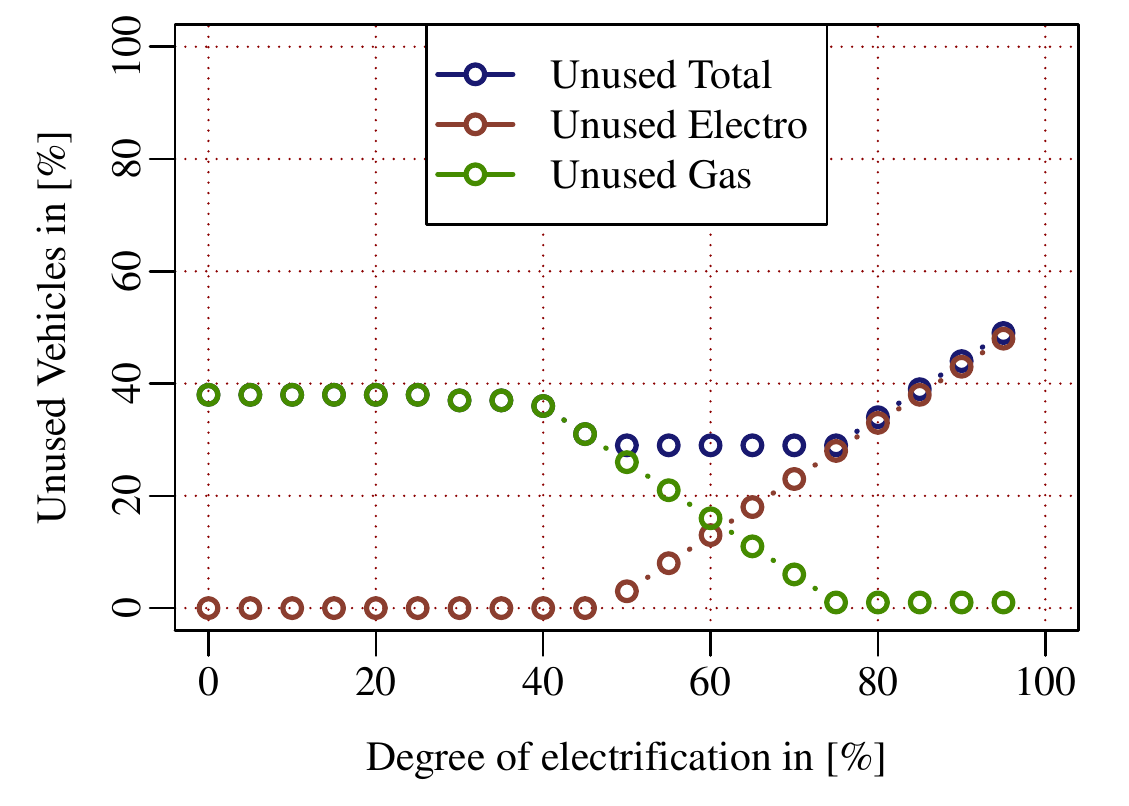}
\caption{Unused vehicles within the scenario}
\label{fig:unused}
\end{figure}

OMNeT++ acts as the control center for our simulation and does not only spawn and assign routes to vehicles but also computes the SOC of the vehicles' batteries and controls the charging and queueing at charging stations and the traffic density in the scenario.
During the simulation, various data is collected from each car for evaluation purposes.
This includes speed and acceleration profiles, the analysis of power flows (consumption, recuperation, range extender), and the investigation of charging and idle periods.
This allows us to draw conclusions as illustrated in \cref{fig:unused}: Considering a certain job schedule and a fixed vehicle fleet size, we are able to identify by how much the fleet is overdimensioned.

%

\section{Communication}

In future work wireless communication will be incorporated in the simulation framework using OMNeT++.
Such simulations will allow the investigation of several aspects such as communicating charging stations exchanging information about 
their occupancy. 
These stations could furthermore coordinate vehicles to achieve some level of load balancing to overcome bottlenecks and reduce waiting times.

Communication between the vehicles and between vehicles and infrastructure could be utilized to lower power consumption as well.
Braking vehicles could send a message to cars behind in order to determine optimal recuperation levels for maximal power recovering.
Communication with the infrastructure, like in Green Light Optimal Speed Advisory (GLOSA) systems \cite{eckhoff2013potentials,tielert2012communication}, could have an impact on the power consumption and thereby on the driving range of electric vehicles, too.

%

\section{Conclusion and Future Work}

In this article we gave an insight in a simulation framework which allows the studying of various aspects of electrified vehicle fleets.
Based on a lightweight battery model, the framework is able to reproduce the power flows within electric cars and to measure the consumption of electric energy.
In addition to that, charging infrastructure has been implemented to reflect recharging of EVs.
This is the basis to, e.g., analyze different layouts of charging infrastructure---considering dimensioning or positioning
of the charging stations.

By having chosen the OMNeT++ simulation framework, we are able to incorporate communication among the different components of our scenario: i.e., cars, charging stations, and vehicle fleet management.



%

\bibliographystyle{IEEEtran}
\bibliography{references}

\end{document}